\documentclass[conference]{IEEEtran}
\usepackage[latin9]{inputenc}
\usepackage[a4paper]{geometry}
\usepackage{array}
\usepackage{amsmath}
\usepackage{amssymb}
\usepackage{graphicx}

\makeatletter

\providecommand{\tabularnewline}{\\}

\@ifundefined{date}{}{\date{}}
\usepackage{cite}

\usepackage{tabularx}
\usepackage{graphicx}

\makeatother

\begin{document}
\title{Outage Analysis of Backscatter-Based Ambient IoT Device Classes with
Energy Buffering }
\author{Azzam Al-nahari$^{\dagger}$, Riku J\"{a}ntti$^{\dagger}$, Yi Zhou$^{\ast}$$^{\mathsection}$,
Deepak Mishra$^{\ddagger}$, Masoud Kaveh$^{\dagger}$, Abhishek Mondal$^{\mathparagraph}$\\
$^{\dagger}${\small{}Department of Information and Communications
Engineering, Aalto University, 02150 Espoo, Finland.}\\
$^{\ast}${\small{}Key Lab of Information Coding and Transmission,
Southwest Jiaotong University, Chengdu, 610031, China.}\\
$^{\mathsection}${\small{}Brunel University London, London, UB8 3PH,
UK }\\
$^{\ddagger}${\small{}School of Electrical Engineering and Telecommunications,
University of New South Wales, Sydney, NSW 2052, Australia.}\\
$^{\mathparagraph}${\small{}National Institute of Technology Calicut,
Kozhikode, Kerala, India}\\
{\small{}Email: \{azzam.al-nahari, riku.jantti, masoud.kaveh\}@aalto.fi,
yizhou@swjtu.edu.cn, d.mishra@unsw.edu.au, abhishekmondal@nitc.ac.in}}
\maketitle
\begin{abstract}
This paper presents an analytical framework for evaluating the outage
probability of ambient Internet of Things (A-IoT) device classes communicating
directly with a base station. Device 1 is a passive backscatter device
with minimal storage, while Device 2 is equipped with a supercapacitor
that enables energy buffering and optional amplification. The proposed
framework jointly accounts for carrier-detection sensitivity, energy
harvesting constraints, supercapacitor energy dynamics, and an energy-aware
amplification policy, while the energy evolution of buffered devices
is modeled using a discrete-time Markov chain (DTMC). The results
highlight the interplay between energy availability and communication
reliability. Device 2 achieves superior performance in energy-rich
regimes due to buffering and amplification gains, whereas Device 1
becomes more robust in energy-constrained regimes, particularly at
larger distances or under high payload requirements. These findings
highlight that the optimal device choice depends critically on the
operating regime and application demands.
\end{abstract}


\section{Introduction}

The rapid expansion of the Internet of Things (IoT) has led to a growing
demand for sustainable, low-cost, and maintenance-free connectivity
for billions of sensors and actuators. As initiated by the 3GPP study
\cite{Tech1}, ambient IoT (A-IoT) has emerged as a promising paradigm
that leverages ambient radio frequency (RF) signals and energy harvesting
to enable batteryless communication \cite{AmbientIoT_Majid}, \cite{Azzam_Mag_2025}.
This approach supports large-scale deployments in applications such
as smart logistics, industrial automation, and environmental monitoring,
where long-term operational autonomy is essential. In the 3GPP Release
19 framework \cite{Tech3}, A-IoT devices are categorized into two
main classes based on hardware complexity and power capability. Device
1 is a passive backscatter device with minimal energy storage, requiring
sufficient harvested energy to meet activation and circuit thresholds.
In contrast, Device 2 is equipped with a supercapacitor that enables
energy buffering across multiple slots, supports higher circuit power,
and can optionally perform reflection amplification to enhance signal
strength. While a version of Device 2 in {[}4{]} may also support
active transmission, this paper focuses on the backscatter-based mode,
where communication is achieved by modulating and reflecting an externally
provided RF carrier.

Backscatter communication has been extensively studied as a low-power
alternative for wireless connectivity \cite{Servey2023}, \cite{27}
and references therein. However, the integration of dedicated energy
harvesting (EH) mechanisms to enable backscatter-based A-IoT remains
a relatively recent and less explored area. Unlike conventional EH-enabled
systems that rely on active transmission \cite{EH2013}, \cite{Mondal2023},
\cite{EH2015}, prior works such as \cite{jameel2019simultaneous,outage4,Gu2023EH,Wang2025EH}
specifically considered EH in backscatter communication. In particular,
\cite{jameel2019simultaneous} analyzed the outage performance under
fading with power splitting, while \cite{outage4} incorporated co-channel
interference and non-linear EH with adaptive reflection control for
minimizing the outage probability. In \cite{Gu2023EH}, an energy
recycling scheme for backscatter communication was proposed, where
IoT devices could harvest energy not only from external sources but
also from the backscatter signals of other devices to improve energy
efficiency. In \cite{Wang2025EH}, an energy harvesting-based CDMA
symbiotic radio system was introduced, where devices harvest energy
from ambient cellular signals and transmit data via backscatter using
distinct spreading codes to enhance throughput. The outage behavior
of different A-IoT connectivity topologies was studied in\cite{Azzam_AIoT2026}. 

Nevertheless, several important gaps remain in the literature. First,
the comparative outage behavior of different A-IoT device classes
under practical energy harvesting constraints has not been systematically
studied. Second, existing works often overlook the joint impact of
carrier-detection sensitivity, stochastic energy harvesting, and energy-aware
amplification on system performance. Third, for buffered devices such
as Device 2, the state-dependent feasibility of operation, including
supercapacitor energy dynamics, leakage, and saturation, has not been
fully characterized within an end-to-end outage probability framework.
Addressing these gaps is essential to understand how performance bottlenecks
shift between energy availability and communication reliability under
varying operating conditions, such as distance and payload requirements.

This paper addresses these gaps by developing an analytical framework
for the outage probability of Device 1 and Device 2, where outage
is jointly determined by energy availability and decoding constraints.
Outage probability is adopted as a suitable performance metric due
to the ultra-low-power operation and inherently intermittent communication
of A-IoT systems. The framework integrates carrier-detection sensitivity
as an activation constraint and employs a discrete-time Markov chain
(DTMC) to model the energy evolution of buffered devices. A time-splitting
protocol is adopted to ensure reliable activation and energy feasibility
prior to communication. For Device 1, a harvesting subslot enables
sufficient energy accumulation to meet circuit requirements within
the same slot. In contrast, Device 2 supports energy accumulation
across multiple slots, allowing the supercapacitor to meet higher
circuit demands and enable optional reflection amplification. The
main contributions of this work are summarized as follows: 
\begin{itemize}
\item We develop a class-specific analytical framework for the outage probability
of A-IoT devices, capturing the distinct operation of Device 1 (in-slot
energy feasibility) and Device 2 (buffered, state-dependent operation),
with carrier-detection sensitivity as an activation constraint. 
\item We model the energy evolution of Device 2 using a DTMC that captures
finite storage, leakage, and state-dependent operation, and use the
resulting stationary distribution to evaluate the outage probability,
incorporating an energy-aware amplification policy. 
\item We analyze the impact of time-splitting and optimize the harvesting
duration, revealing the interplay between energy availability and
communication reliability and identifying the operating regimes of
each device class.
\end{itemize}

\section{System Model}

We consider a monostatic A-IoT topology, as illustrated in Fig. 1,
where an indoor base station (BS) provides the RF carrier. The\,A\nobreakdash-IoT
device communicates directly with the\,BS by modulating and reflecting
the incident carrier signal, consistent with\,Topology\,1 in\, \cite{Tech3}.
The BS thus acts as the carrier emitter and reader, transmitting an
unmodulated RF carrier and receiving the backscattered signal. All
nodes are equipped with a single antenna. A time-splitting protocol
is adopted, where each slot $T$ consists of an energy harvesting
phase $T_{h,k}=\tau_{k}T$ followed by a backscatter phase $T_{b,k}=(1-\tau_{k})T$,
where $k\in\{1,2\}$ denotes the device index and $\tau_{k}$ is the
time allocation factor.

\subsection{Carrier Reception and Device Wake-up}

\begin{figure}[t]
\begin{centering}
\includegraphics[width=7cm,height=6cm]{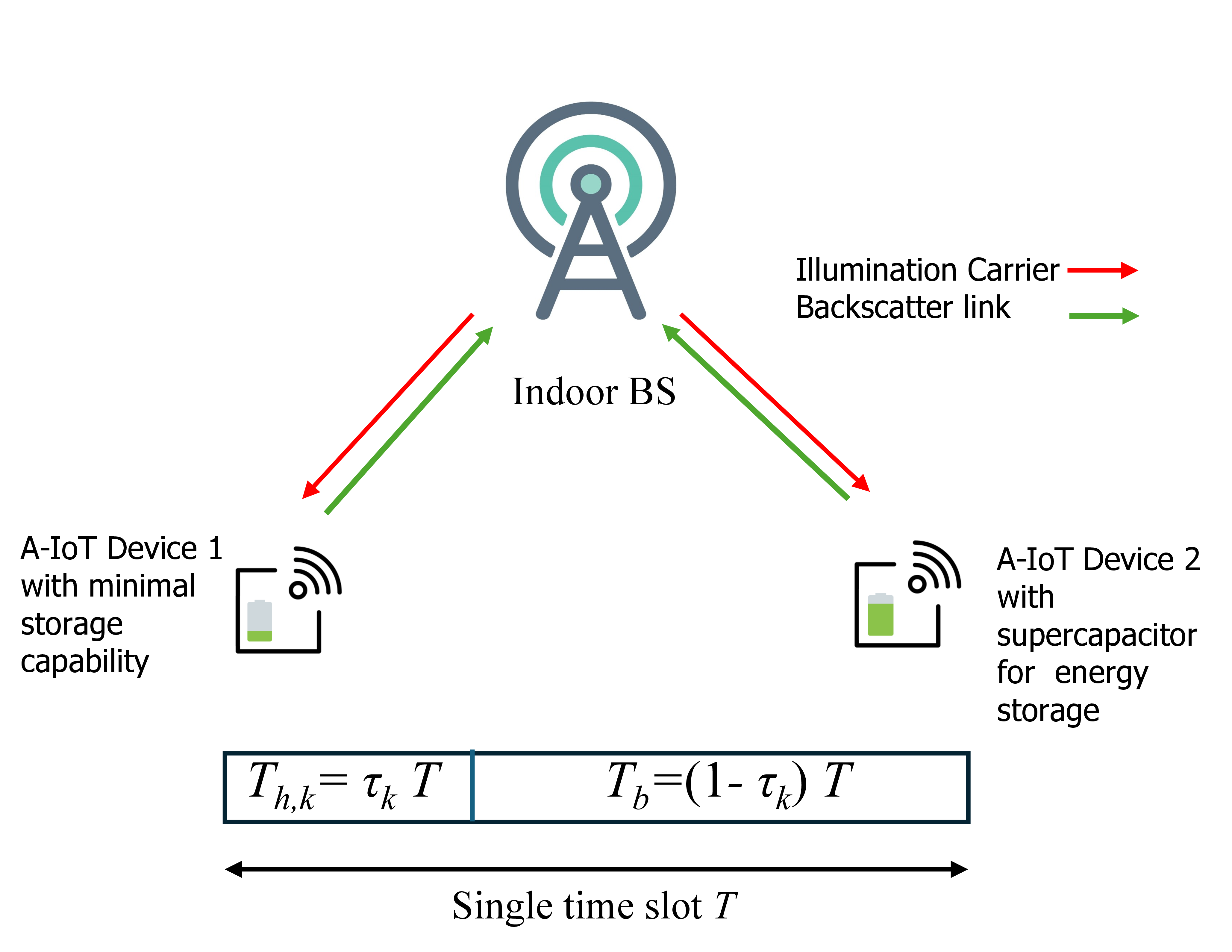}
\par\end{centering}
\centering{}\caption{System model of an A-IoT network with an indoor BS and two distinct
A-IoT device classes (Device 1 and Device 2).}
\end{figure}

\begin{table}
\caption{List of common symbols.}

\centering{}%
\begin{tabular}{|>{\centering}p{1cm}|>{\raggedright}p{6.1cm}|}
\hline 
Symbol & \centering Description\tabularnewline
\hline 
\hline 
$E_{h,n}$ & harvested energy at the device during slot $n$\tabularnewline
\hline 
$E_{\mathrm{\mathit{c},\mathit{k}}}$ & required energy for Device $k$ operation in one slot\tabularnewline
\hline 
$E_{\mathrm{amp}}$  & additional energy cost to enable amplification in a slot (Device~2) \tabularnewline
\hline 
$E_{n}$ & stored energy in Device~2 supercapacitor at the \emph{start} of slot
$n$\tabularnewline
\hline 
$E_{\max}$  & maximum energy capacity of the supercapacitor\tabularnewline
\hline 
$P_{c,k}$ & peak/active circuit power consumption of Device~$k$\tabularnewline
\hline 
$P_{\mathrm{cd,\mathit{k}}}$ & carrier-detection sensitivity of device $k$ \tabularnewline
\hline 
$P_{\mathrm{amp}}$ & amplification power for Device 2\tabularnewline
\hline 
$P_{I}$ & BS illumination carrier transmit power\tabularnewline
\hline 
$G_{a}$ & Effective SNR gain of Device 2 due to amplification\tabularnewline
\hline 
$\rho\in(0,1]$ & the retention factor per slot, capturing supercapacitor leakage\tabularnewline
\hline 
$\eta_{k}$ & RF-to-DC energy conversion efficiency for Device $k$\tabularnewline
\hline 
$\alpha_{k}$ & reflection coefficient of device $k$\tabularnewline
\hline 
\end{tabular}
\end{table}
The instantaneous received RF power at Device $k$ from the BS in
time slot $n$ is given by 
\begin{equation}
P_{r,k,n}=P_{I}\,|h_{b,k}|^{2},
\end{equation}
where $P_{I}$ denotes the transmit power of the BS illumination carrier,
and $h_{b,k}\sim\mathcal{CN}(0,\beta_{k})$ represents the fading
channel between the BS and device $k$. The average channel gain is
$\beta_{k}=d_{k}^{-\vartheta}$, where $d_{k}$ is the distance between
the BS and Device $k$, and $\vartheta=3$ is the path loss exponent.
We assume block fading with independent channel realizations across
slots, leading to independent harvested energy realizations and enabling
a DTMC-based energy model. The device operates (i.e., detects the
carrier and wakes up) only if the received power exceeds its carrier-detection
sensitivity $P_{\mathrm{cd,\mathit{k}}}$. Accordingly, the \textit{device-activation
event} of device $k$ in time slot $n$ is defined as
\begin{equation}
\mathcal{A}_{k,n}\triangleq\{P_{r,k,n}\ge P_{\mathrm{cd,\mathit{k}}}\}.
\end{equation}

\subsection{Energy Harvesting Model }

The device harvests energy from the incident carrier. A linear harvesting
model with a detection threshold is adopted: if the device cannot
detect the carrier (i.e., $\mathcal{A}_{k,n}$ is false), the harvested
energy is zero. Accordingly, the harvested energy in time slot $n$
is given by
\begin{equation}
E_{h,k,n}=\begin{cases}
\eta_{k}\,P_{r,k,n}\,T_{h,k}, & \text{if }\mathcal{A}_{k,n}\text{ holds},\\
0, & \text{if }\mathcal{A}_{k,n}\text{ does not hold}.
\end{cases}
\end{equation}
where $\eta_{k}\in(0,1]$ is the RF-to-DC energy conversion efficiency.
Substituting (1) into (3), when $\mathcal{A}_{k,n}$ holds, yields
\begin{equation}
E_{h,k,n}=\eta_{k}\,P_{I}\,|h_{b,k}|^{2}\,T_{h,k}.
\end{equation}

\subsection{Device 1: Harvest-Then-Transmit Without Buffering}

Device 1 has minimal storage and operates in a harvest-then-transmit
manner within each time slot, with no meaningful energy carry-over
across slots. The required circuit energy consumption during the backscatter
phase is given by
\begin{equation}
E_{c,1}=P_{c,1}\,T_{b,1},
\end{equation}
where $P_{c,1}$ denotes the active circuit power, and $T_{b,1}$
is the backscatter duration. The \emph{energy-sufficient event} is
defined as
\begin{equation}
\mathcal{E}_{1,n}\triangleq\{E_{h,1,n}\ge E_{c,1}\}.
\end{equation}
The \emph{feasibility event}, requiring both carrier detection and
sufficient harvested energy within slot $n$, is given by
\begin{equation}
\mathcal{F}_{1,n}\triangleq\mathcal{A}_{1,n}\cap\mathcal{E}_{1,n}.
\end{equation}

\subsection{Device 2: Harvest-then-Transmit With Energy Buffer and Amplification}

Device~2 employs a supercapacitor to buffer energy across slots.
Let $E_{n}\in[0,E_{\max}]$ denote the stored energy at the start
of slot $n$, where $E_{\max}$ is the supercapacitor capacity. The
circuit energy consumption per slot is given by 
\begin{equation}
E_{c,2}=P_{c,2}\,T_{b,2},
\end{equation}
where $P_{c,2}$ is the active circuit power. The energy-sufficient
event for Device~2, ensuring that enough energy is available to operate
the circuitry in slot $n$, is defined as
\begin{equation}
\mathcal{E}_{2,n}^{\mathrm{pas}}\triangleq\{\rho E_{n}+E_{h,2,n}\ge E_{c,2}\}.
\end{equation}
where $\rho\in(0,1]$ denotes the energy retention factor accounting
for leakage. The feasibility event is then given by

\begin{equation}
\mathcal{F}_{2,n}\triangleq\mathcal{A}_{2,n}\cap\mathcal{E}_{2,n}^{\mathrm{pas}}.
\end{equation}

\subsubsection{Energy-Aware Amplification Policy}

Device 2 incurs additional energy consumption $E_{\mathrm{amp}}=T_{b,2}P_{\mathrm{amp}}$
when amplification is enabled in slot $n$, where $P_{\mathrm{amp}}$
is the amplification power. Accordingly, the energy-sufficient event
for amplified operation is defined as

\begin{equation}
\mathcal{E}_{2,n}^{\mathrm{amp}}\triangleq\{\rho E_{n}+E_{h,2,n}\ge E_{c,2}+E_{\mathrm{amp}}\}.
\end{equation}
We define $u_{n}\triangleq\mathbf{1}\{\mathcal{E}_{2,n}^{\mathrm{amp}}\}$,
where $\mathbf{1}\{\cdot\}$ denotes the indicator function, which
equals 1 if the condition inside the braces is satisfied and 0 otherwise.
Accordingly, $u_{n}=1$ indicates that amplification is enabled, incurring
an additional energy cost $E_{\mathrm{amp}}$ and providing an SNR
gain $G_{a}$, while $u_{n}=0$ corresponds to passive operation with
no additional energy cost or SNR gain.

\subsubsection{Supercapacitor Energy Evolution}

The energy state of the supercapacitor evolves as 
\begin{equation}
E_{n+1}=\min\Big(E_{\max},\,[\rho E_{n}+E_{h,2,n}-a_{n}(E_{c,2}+u_{n}E_{\mathrm{amp}})]^{+}\Big),
\end{equation}
where $a_{n}\triangleq\mathbf{1}\{\mathcal{A}_{2,n}\}\,\mathbf{1}\{\mathcal{E}_{2,n}^{\mathrm{pas}}\}$,
$[x]^{+}\triangleq\max[x,0]$. The stored energy evolves across time
slots based on the balance between retained energy, harvested energy,
and consumption. In (12), we model leakage by scaling the stored energy
at the start of each slot by\,$\rho$, since the supercapacitor\textquoteright s
self\nobreakdash-discharge is slow relative to the slot duration.
Thus, the energy harvested during the slot is not reduced by leakage
until the next slot, when the entire stored energy---including any
unspent harvested energy---is scaled by\,$\rho$ again. Energy is
consumed only when the device is active, as indicated by $a_{n}$.
In this case, the device expends energy $E_{c,2}$, and additional
energy $E_{\mathrm{amp}}$ if amplification is enabled, as determined
by $u_{n}$.\\
\\
\\

\subsection{Signal Model}

During the backscatter phase, the A-IoT device modulates and reflects
the incident carrier toward the BS. The backscattered signal from
Device $k$ at the BS is given by

\begin{equation}
y_{b,k}=\sqrt{\alpha_{k}P_{I}}h_{b,k}h_{k,b}x_{k}+n_{b},
\end{equation}
where $x_{k}$ is the information signal backscattered from Device
$k$ with $\mathbb{E}[|x_{k}|^{2}]=1$, $\alpha_{k}$ denotes the
reflection coefficient, and $n_{b}\mathcal{\sim CN}(0,N_{0})$ is
the additive white Gaussian noise (AWGN). Assuming symmetric fading
channels, i.e., $h_{b,k}=h_{k,b}$, the SNR at the BS for Device 1
is given by 
\begin{equation}
\gamma_{1,n}=\frac{\alpha_{1}P_{I}\,|h_{b,1}|^{4}}{N_{0}}.
\end{equation}
For Device~2, the SNR depends on the amplification status and is
given by

\begin{equation}
\gamma_{2,n}=\begin{cases}
\frac{\alpha_{2}P_{I}\,|h_{b,2}|^{4}}{N_{0}}\triangleq\gamma_{2,n}^{\mathrm{pas}}, & \text{if }u_{n}=0,\\
\frac{G_{a}\alpha_{2}P_{I}\,|h_{b,2}|^{4}}{N_{0}}\triangleq\gamma_{2,n}^{\mathrm{amp}}, & \text{if }u_{n}=1.
\end{cases}
\end{equation}
where $G_{a}$ denotes the effective SNR gain achieved through active
reflection amplification, with amplifier noise neglected relative
to the BS noise. The achievable rate for Device $k$ is given as $R_{k,n}=\log_{2}(1+\gamma_{k,n})$.
Accordingly, the \textit{decoding-success event} at the BS for Device
$k$ is defined as 

\begin{equation}
\mathcal{S}_{k,n}\triangleq\{\gamma_{k,n}\ge2^{\frac{B}{WT_{b,k}}}-1\},
\end{equation}
where $\frac{B}{WT_{b,k}}$ (in bits per channel use) is the target
rate, $B$ is the number of transmitted bits during the backscatter
phase, and $W$ is the channel bandwidth. 

\section{Outage Probability Analysis }

End-to-end success in slot $n$ requires that the device detects the
carrier ($\mathcal{A}_{k,n}$), satisfies the energy constraint ($\mathcal{E}_{1,n}$
for Device 1, and $\mathcal{E}_{2,n}^{\mathrm{pas}}/\mathcal{E}_{2,n}^{\mathrm{amp}}$
for Device 2), and that the BS successfully decodes the backscattered
signal ($\mathcal{S}_{k,n}$).\\

\subsection{Outage Performance for Device 1 }

The end-to-end success event for Device~1 is defined as
\begin{equation}
\mathcal{G}_{1,n}\triangleq\mathcal{A}_{1,n}\cap\mathcal{E}_{1,n}\cap\mathcal{S}_{1,n}.
\end{equation}
Thus, the outage probability is 
\begin{align}
P_{\mathrm{out},1} & =1-\Pr(\mathcal{G}_{1,n})\nonumber \\
 & =1-\Pr(\mathcal{A}_{1,n}\cap\mathcal{E}_{1,n}\cap\mathcal{S}_{1,n})\nonumber \\
 & =1-\Pr\left(|h_{b,1}|^{2}\ge\theta_{1},|h_{b,1}|^{2}\ge\theta_{2},|h_{b,1}|^{2}\ge\theta_{3}\right)
\end{align}
where $\theta_{1}\triangleq\frac{P_{\mathrm{cd,\mathit{\mathrm{1}}}}}{P_{I}}$,
$\theta_{2}\triangleq\frac{P_{c,1}(1-\tau_{1})}{\eta_{1}\,P_{I}\tau_{1}}$,
$\theta_{3}\triangleq\sqrt{\frac{(2^{\frac{B}{W(1-\tau_{1})T}}-1)N_{0}}{\alpha_{1}P_{I}}}$
correspond to detection, energy, and decoding constraints, respectively.
As $\bigl|h_{b,1}\bigr|^{2}$ is exponentially distributed with parameter
$\frac{1}{\beta_{1}}$, the outage probability is given by

\begin{equation}
P_{\mathrm{out},1}=1-\exp\bigl(-\max(\theta_{1},\theta_{2},\theta_{3})/\beta_{1}\bigr)
\end{equation}

\subsection{Outage Performance for Device 2}

Device~2 operates in two modes depending on whether amplification
is enabled.

\subsubsection{Amplified Mode Success Event}

The end-to-end success event in the amplified mode is defined as

\begin{equation}
\mathcal{G}_{2,n}^{\mathrm{amp}}\triangleq\mathcal{A}_{2,n}\cap\mathcal{E}_{2,n}^{\mathrm{amp}}\cap\{\gamma_{2,n}^{\mathrm{amp}}\ge2^{\frac{B}{WT_{b,2}}}-1\}.
\end{equation}
Since this event depends on the stored energy $E_{n}$, which evolves
stochastically, the success probability is evaluated conditionally
on a given energy state $E_{n}=s_{j}$ and later averaged over its
stationary distribution (see Appendix A). Conditioned on $E_{n}=s_{j}$,
the success probability is given by

\begin{align}
\Pr(\mathcal{G}_{2,n}^{\mathrm{amp}}|s_{j}) & =\Pr\left(|h_{b,2}|^{2}\ge\theta_{4},|h_{b,2}|^{2}\ge\theta_{5},|h_{b,2}|^{2}\ge\theta_{6}\right)\nonumber \\
 & =\exp\Bigl(-\frac{\max(\theta_{4},\theta_{5},\theta_{6})}{\beta_{2}}\Bigr)
\end{align}
where $\theta_{4}\triangleq\frac{P_{\mathrm{cd,\mathit{\mathrm{2}}}}}{P_{I}}$,
$\theta_{5}\triangleq\frac{(P_{c,2}+P_{\mathrm{amp}})(1-\tau_{2})}{\eta_{2}\,P_{I}\tau_{2}}-\frac{\rho s_{j}}{\eta_{2}\,P_{I}\tau_{2}T}$,
and $\theta_{6}\triangleq\sqrt{\frac{(2^{\frac{B}{W(1-\tau_{2})T}}-1)N_{0}}{G_{a}\alpha_{2}P_{I}}}$
.

\subsubsection{Passive Mode Success Event}

In the passive mode, the end-to-end success event is defined as

\begin{align}
\mathcal{G}_{2,n}^{\mathrm{pas}}\triangleq & \mathcal{A}_{2,n}\cap\mathcal{E}_{2,n}^{\mathrm{pas}}\cap\{\rho E_{n}+E_{h,2,n}<E_{c,2}+E_{\mathrm{amp}}\}\nonumber \\
 & \cap\{\gamma_{2,n}^{\mathrm{pas}}\ge2^{\frac{B}{WT_{b,2}}}-1\}.
\end{align}
Here, the condition $E_{n}+E_{h,2,n}<E_{c,2}+E_{\mathrm{amp}}$ excludes
the amplification regime. As in the amplified case, the success probability
is evaluated conditioned on $E_{n}=s_{j}$ as

\begin{align}
\Pr(\mathcal{G}_{2,n}^{\mathrm{pas}}|s_{j}) & =\Pr\Bigl(|h_{b,2}|^{2}\ge\theta_{4},|h_{b,2}|^{2}<\theta_{5},\nonumber \\
 & \,\,\,\,\,\,\,\,\,\,\,|h_{b,2}|^{2}\ge\theta_{7},|h_{b,2}|^{2}\ge\theta_{8}\Bigr)\nonumber \\
 & =F_{|h_{b,2}|^{2}}(\theta_{5})-F_{|h_{b,2}|^{2}}(\max(\theta_{4},\theta_{7},\theta_{8}))\nonumber \\
 & =v_{j}^{\mathrm{pas}}\Bigl[\exp\Bigl(-\frac{\max(\theta_{4},\theta_{7},\theta_{8})}{\beta_{2}}\Bigr)-\exp\Bigl(-\frac{\theta_{5}}{\beta_{2}}\Bigr)\Bigr]
\end{align}
where $\theta_{7}\triangleq\frac{P_{c,2}(1-\tau_{2})}{\eta_{2}\,P_{I}\tau_{2}}-\frac{\rho s_{j}}{\eta_{2}\,P_{I}\tau_{2}T}$,
and $\theta_{8}\triangleq\sqrt{\frac{(2^{\frac{B}{W(1-\tau_{2})T}}-1)N_{0}}{\alpha_{2}P_{I}}}$.
The indicator function $v_{j}^{\mathrm{pas}}\triangleq\mathbf{1}\{\max(\theta_{4},\theta_{7},\theta_{8})<\theta_{5}\}$
in (23) enforces the condition $\max(\theta_{4},\theta_{7},\theta_{8})<\theta_{5}$;
if this condition is not met the passive\nobreakdash-mode interval
is empty and the passive probability term is set to zero. Since $\mathcal{G}_{2,n}^{\mathrm{amp}}$and
$\mathcal{G}_{2,n}^{\mathrm{pas}}$ are disjoint, the conditional
outage probability is 
\begin{align}
P_{\mathrm{out},2|s_{j}} & =1-\Pr(\mathcal{G}_{2,n}^{\mathrm{amp}}|s_{j})-\Pr(\mathcal{G}_{2,n}^{\mathrm{pas}}|s_{j}).
\end{align}
Since the stored energy in (12) evolves over time, it is modeled as
a finite $K$-state DTMC to enable tractable steady-state analysis
via the stationary distribution $\pi_{j}$ of the state $s_{j}$ (Appendix~A).
Accordingly, the overall outage probability is

\begin{align}
P_{\mathrm{out},2} & =1-\sum_{j=0}^{K-1}\pi_{j}\left(\Pr(\mathcal{G}_{2,n}^{\mathrm{amp}}\mid s_{j})+\Pr(\mathcal{G}_{2,n}^{\mathrm{pas}}\mid s_{j})\right)\nonumber \\
 & =1-\sum_{j=0}^{K-1}\pi_{j}\biggl[\exp\Bigl(-\frac{\max(\theta_{4},\theta_{5},\theta_{6})}{\beta_{2}}\Bigr)\nonumber \\
 & \,\,\,\,\,\,\,\,\,\,+v_{j}^{\mathrm{pas}}\Bigl[\exp\Bigl(-\frac{\max(\theta_{4},\theta_{7},\theta_{8})}{\beta_{2}}\Bigr)-\exp\Bigl(-\frac{\theta_{5}}{\beta_{2}}\Bigr)\Bigr]\biggr]
\end{align}

\subsection{Optimization of $\tau_{k}$}

In this subsection, we determine the optimal value of $\tau_{k}$.
The optimal harvesting fraction is obtained by solving 
\begin{equation}
\tau_{k}^{\star}=\arg\min_{0<\tau_{k}<1}P_{\mathrm{out},k}(\tau_{k}).
\end{equation}

\subsubsection{Device 1}

Since $P_{\mathrm{out},1}$ in (19) is monotonic in $\max(\theta_{1},\theta_{2},\theta_{3})$,
minimizing $P_{\mathrm{out},1}$ is equivalent to solving 
\begin{equation}
\min_{0<\tau_{1}<1}\;\max\!\left(\theta_{1},\theta_{2}(\tau_{1}),\theta_{3}(\tau_{1})\right).
\end{equation}
It is observed that $\theta_{2}(\tau_{1})$ is monotonically decreasing
in $\tau_{1}$, whereas $\theta_{3}(\tau_{1})$ is monotonically increasing.
Hence, the optimal time allocation is achieved when the two competing
thresholds are balanced, i.e., $\theta_{2}(\tau_{1}^{\star})=\theta_{3}(\tau_{1}^{\star})$.
Accordingly, $\tau_{1}^{\star}$ is determined as the root of $f(\tau_{1})=\theta_{2}(\tau_{1})-\theta_{3}(\tau_{1})=0$.
Thus, the optimal time allocation factor is given by 
\begin{equation}
\tau_{1}^{\star}=\left\{ \tau_{1}\in(0,1):f(\tau_{1})=0\right\} ,
\end{equation}
which can be efficiently obtained via a one-dimensional search (e.g.,
using the \texttt{fzero} function in MATLAB). When $\theta_{1}$ dominates
in (27), the objective is independent of $\tau_{1}$.

\subsubsection{Device 2}

The outage probability in (25) depends on $\tau_{2}$ through multiple
terms (energy thresholds, decoding constraint, and stationary distribution
$\mathbf{\pi}$). Unlike Device~1, no closed-form solution exists.
Thus, $\tau_{2}^{\star}$ is obtained numerically using (26) by finding
the transition matrix, computing $\mathbf{\pi}$, and evaluating $P_{\mathrm{out},2}(\tau_{2})$,
followed by a one-dimensional search.

\section{Numerical Results }

The simulation parameters are selected to align with the assumptions
outlined in \cite{Tech3}. Unless otherwise stated, we set ~$P_{c,1}=1\mu$W
and $P_{c,2}=200\,\mu$W, $W=15$~kHz, $B=200\,\mathrm{bits}$, $\alpha_{1}=\alpha_{2}=0.5$,
$\eta_{1}=\eta_{2}=0.5$, $P_{\mathrm{cd},1}=-30\,\mathrm{dBm}$,
$P_{\mathrm{cd},2}=-40\,\mathrm{dBm}$, $T=5\,\mathrm{ms}$, $d_{k}=10\,\mathrm{m}$,
and $\rho=0.95$. A nominal capacitance of 68~$\mu$F at $2\,\mathrm{V}$
is adopted for Device 2, yielding a maximum stored energy of $E_{\max}=\tfrac{1}{2}CV^{2}\approx136~\mu$J.
The reflection amplifier gain of Device 2 is set to $G_{a}=15\,\mathrm{dB}$,
with a corresponding power consumption of $P_{\mathrm{amp}}=40\mu\mathrm{W}$. 

\begin{figure}[t]
\begin{centering}
\includegraphics[clip,width=8cm,height=4.6cm]{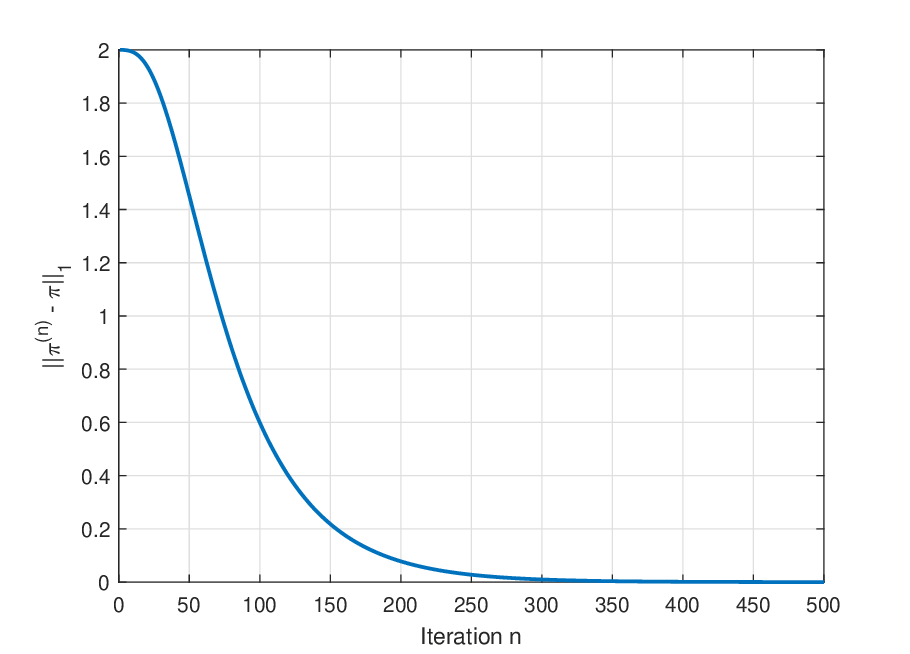}
\par\end{centering}
\centering{}\caption{Convergence of the state probability vector $(K=300)$.}
\end{figure}

\begin{figure}[t]
\begin{centering}
\includegraphics[clip,width=8cm,height=4.6cm]{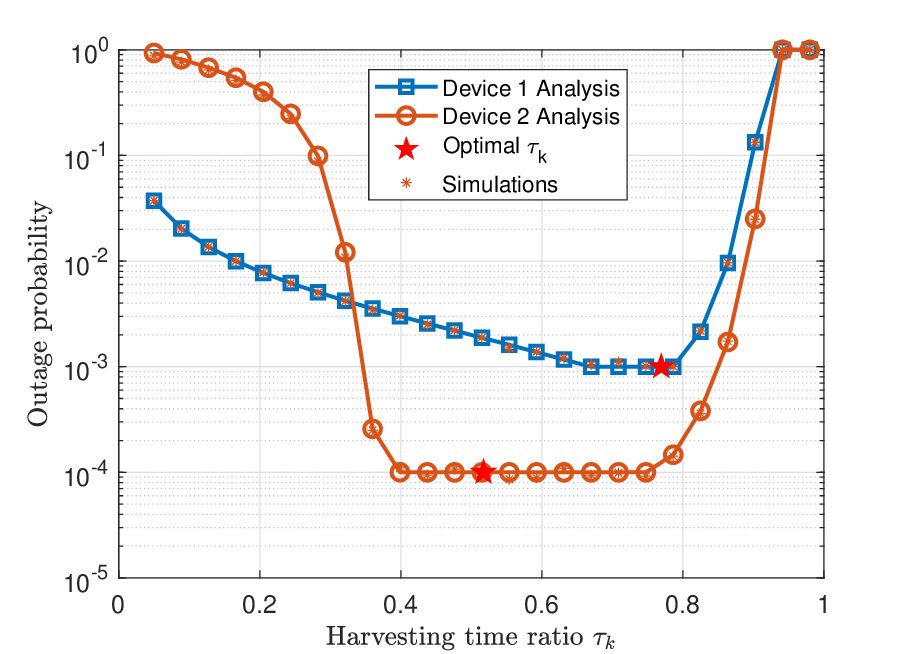}
\par\end{centering}
\centering{}\caption{Outage probability versus harvesting time ratio $\tau_{k}$.}
\end{figure}

\begin{figure}[t]
\begin{centering}
\includegraphics[clip,width=8cm,height=4.6cm]{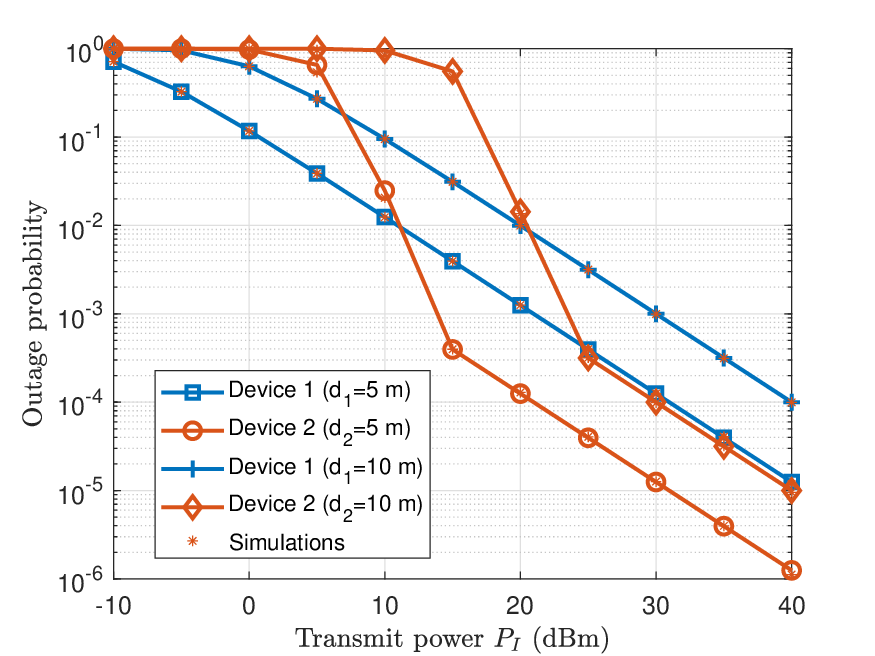}
\par\end{centering}
\centering{}\caption{Outage probability versus transmit power $P_{I}$.}
\end{figure}

\begin{figure}[t]
\begin{centering}
\includegraphics[clip,width=8cm,height=4.7cm]{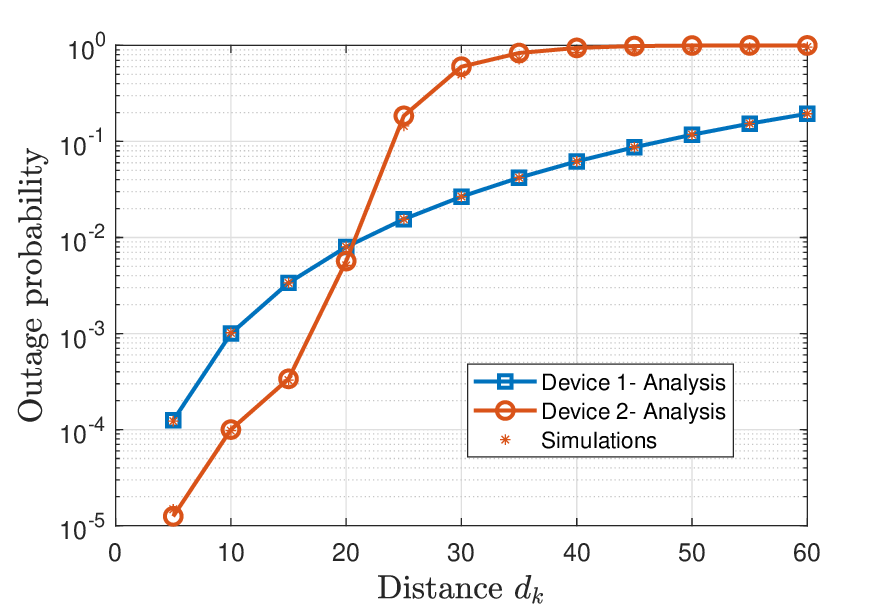}
\par\end{centering}
\centering{}\caption{Outage probability versus distance $d_{k}$ under optimal $\tau_{k}^{\star}$.}
\end{figure}

\begin{figure}[t]
\begin{centering}
\includegraphics[clip,width=8cm,height=4.5cm]{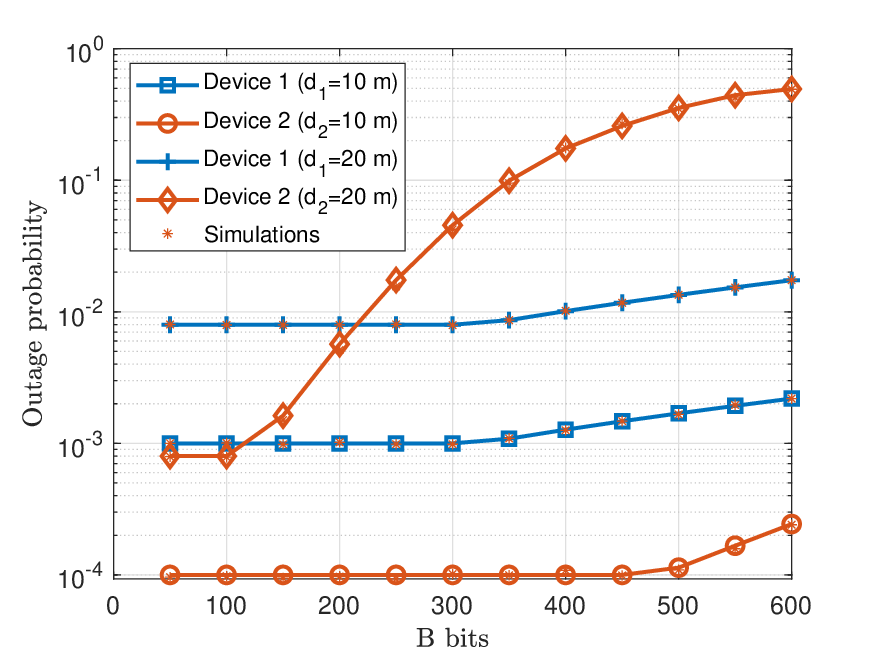}
\par\end{centering}
\centering{}\caption{Outage probability versus payload $B$ (bits per time slot).}
\end{figure}

Fig. 2 illustrates the convergence of the Markov process by showing
the norm of the difference between successive probability distributions.
The norm decreases monotonically to zero, indicating convergence to
a steady-state distribution and confirming the stability of the proposed
model.

Fig.~3 illustrates the outage probability as a function of $\tau_{k}$.
For small harvesting durations, Device~2 exhibits higher outage due
to insufficient harvested energy to meet its higher circuit requirements,
resulting in energy-limited operation. As $\tau_{k}$ increases, sufficient
energy enables circuit operation and amplification, leading to a sharp
reduction in outage. For large $\tau_{k}$, the outage of both devices
saturates, indicating a transition to a communication-limited regime.
In this regime, reduced backscatter time $T_{b,k}$ increases the
required rate $2^{\frac{B}{WT_{b,k}}}$, and performance becomes dominated
by fading. This reveals an optimal operating point balancing energy
availability and communication reliability due to reduced transmission
time.

Fig. 4 shows the outage probability versus transmit power. At low
and moderate power levels, Device 2 exhibits higher outage probability
than Device 1 due to its larger circuit and amplification energy requirements,
resulting in energy-limited operation. As the transmit power increases,
a sharp transition is observed, where sufficient harvested energy
enables reliable operation and amplification. This leads to a rapid
reduction in outage probability for Device 2, which then outperforms
Device 1 at high power levels due to its buffering and amplification
gains.

Fig.~5 shows the outage probability versus distance under optimal
time allocation. As distance increases, path loss reduces the received
power, leading to lower harvested energy and SNR, and thus higher
outage for both devices. Device~2 outperforms Device~1 at short
distances by leveraging buffered energy for active amplification,
improving its effective SNR. However, at larger distances, its higher
circuit and energy demands lead to frequent outages. In contrast,
Device~1 remains more robust due to its ultra-low power consumption,
outperforming Device~2 in energy-limited regimes. This reveals a
distance threshold beyond which the added complexity of Device~2
becomes detrimental under RF energy harvesting constraints, indicating
that simpler architectures are more suitable for long-range, low-power
operation in such scenarios.

Fig.~6 shows the outage probability versus the payload $B$ for both
devices at different distances. At $d_{k}=10\,\mathrm{m}$, Device~2
outperforms Device~1 due to sufficient harvested energy enabling
amplification and higher SNR. A similar trend holds at $d_{k}=20\,\mathrm{m}$
for low to moderate $B$. As $B$ increases, the system becomes energy-limited,
particularly at larger distances, where Device~2 lacks sufficient
energy for its higher circuit and amplification demands, leading to
frequent outages, while Device~1 remains more efficient. Consequently,
Device~1 outperforms Device~2 at high payloads and longer distances.
This indicates that Device~2 is better suited for high payload transmission
near the BS, while Device~1 is more robust for larger payloads and
longer ranges. 

\section{Conclusion}

This paper develops an analytical framework for the outage probability
of A-IoT devices under RF energy harvesting, capturing the distinct
operation of Device 1 (without buffering) and Device 2 (with buffering
and amplification). The model incorporates carrier-detection sensitivity,
energy availability, and energy-aware operation, enabling a unified
analysis of energy and decoding limitations. Results show that Device
2 outperforms Device 1 in energy-rich conditions, while under energy-limited
regimes (e.g., large distances or low incident power), its higher
energy consumption leads to frequent outages, especially at high payloads.
In contrast, Device 1 is more robust due to its lower energy demands.
Overall, performance depends on the interplay between energy availability
and communication requirements, providing insights for A-IoT device
design and standardization.

\appendices{}

\section{Markov Chain Analysis for Device 2 Outage Probability}

To evaluate the stationary outage probability of Device 2 given in
(25), we discretize the supercapacitor energy $E_{n}\in[0,E_{max}]$
into $K$ discrete levels. Let the state space be $\mathcal{S}=\{s_{0},s_{1},\dots,s_{K-1}\}$,
where $s_{k}=k\Delta$ and $\Delta=\frac{E_{\mathrm{max}}}{K-1}$.

\subsection{State Transition Probabilities}

Recall from (12) that, given $E_{n}=s_{j}$, the energy state at the
beginning of slot $n+1$ evolves as
\begin{equation}
E_{n+1}=\min(E_{\mathrm{max}},[\rho s_{j}+E_{h,2,n}-a_{n}(E_{c,2}+u_{n}E_{\mathrm{amp}})]^{+})
\end{equation}
The transition probability from state $s_{j}$ to $s_{k}$ is defined
as
\begin{equation}
P_{j,k}=\Pr\left((k-0.5)\Delta\le E_{n+1}\le(k+0.5)\Delta\mid E_{n}=s_{j}\right)
\end{equation}
To capture the effect of the wireless channel on the energy evolution,
$P_{j,k}$ is expressed in terms of the small-scale fading gain $|h_{b,2}|^{2}$.
Define $C_{j}=a_{n}(E_{c,2}+u_{n}E_{\mathrm{amp}})$ as the energy
consumption cost in state $s_{j}$. Substituting the energy evolution
model into (30), the transition from $s_{j}$ to $s_{k}$ occurs if 

\begin{equation}
(k-0.5)\Delta\le(\rho s_{j}+\eta_{2}\,P_{I}\,|h_{b,2}|^{2}\,T_{h,2}-C_{j})\le(k+0.5)\Delta
\end{equation}
Let $\Omega=\eta_{2}P_{I}T_{h,2}$ denote the harvesting scaling factor.
Solving (31) for the random variable $X\triangleq|h_{b,2}|^{2}$,
the lower and upper thresholds are 
\begin{align}
L_{j,k} & =\max\Bigl(0,\frac{1}{\Omega}\left[(k-0.5)\Delta+C_{j}-\rho s_{j}\right]\Bigr)\\
U_{j,k} & =\max\Bigl(0,\frac{1}{\Omega}\left[(k+0.5)\Delta+C_{j}-\rho s_{j}\right]\Bigr)
\end{align}
As $X$ follows an exponential distribution, we get 
\begin{equation}
P_{j,k}=\int_{L_{j,k}}^{U_{j,k}}\frac{1}{\beta_{2}}e^{-\frac{x}{\beta_{2}}}dx=e^{-\frac{L_{j,k}}{\beta_{2}}}-e^{-\frac{U_{j,k}}{\beta_{2}}}
\end{equation}
Due to the state-dependent activation and amplification policy, the
transition probability is inherently piecewise in $X$. The channel
gain domain can be partitioned into regions corresponding to different
operating modes, each yielding distinct bounds in (34). For a given
state $E_{n}=s_{j}$, define $x_{\mathrm{pas}}=\max\Bigl(\theta_{4},\Bigl[\frac{E_{c,2}-\rho s_{j}}{\Omega}\Bigr]^{+}\Bigr)$,
and $x_{\mathrm{amp}}=\max\Bigl(\theta_{4},\Bigl[\frac{E_{c,2}+E_{\mathrm{amp}}-\rho s_{j}}{\Omega}\Bigr]^{+}\Bigr)$.
Then, the four channel-gain regions are $0\leq X<\theta_{4}$ for
no activation, where no energy is harvested and no energy is consumed;
$\theta_{4}\leq X<x_{\mathrm{pas}}$ for activation but insufficient
energy for backscatter, where $C_{j}=0$; $x_{\mathrm{pas}}\leq X<x_{\mathrm{amp}}$
for passive backscatter operation, where $C_{j}=E_{c,2}$; and $X\geqslant x_{\mathrm{amp}}$
for amplified backscatter operation, where $C_{j}=E_{c,2}+E_{\mathrm{amp}}$.
The overall transition probability in (34) is obtained by summing
these region-wise integrations. For brevity, the detailed derivation
is omitted.

\subsection{Steady-State Distribution}

The transition matrix $\mathbf{P}\in\mathbb{R}^{K\times K}$ contains
the elements $P_{j,k}$. The stationary distribution vector $\boldsymbol{\pi}=[\pi_{0},\dots,\pi_{K-1}]$
represents the probability that the device is in state $s_{j}$ at
the beginning of a slot. It is obtained by solving

\begin{equation}
\boldsymbol{\pi}\mathbf{P}=\boldsymbol{\pi},\quad\text{subject to }\sum_{j=0}^{K-1}\pi_{j}=1
\end{equation}
Finally, the steady-state probabilities $\pi_{j}$ are used in (25)
to compute the overall outage probability.

\section*{Acknowledgment}

The work of A. Al-nahari, R. J\"{a}ntti, and M. Kaveh was supported in
part by the Smart Networks and Services Joint Undertaking under the
EU\textquoteright s Horizon Europe Research and Innovation Programme
under Grant Agreement No. 101192113 (Ambient-6G). D. Mishra's participation
was supported in part by the Australian Research Council Discovery
Early Career Researcher Award (DECRA) - DE230101391. A. Mondal\textquoteright s
participation was supported in part by the Anusandhan National Research
Foundation, Government of India, through Grant ANRF/IRG/2024/000503/ENS-G,
and by NIT Calicut through Grant NITC/PRJ/ECED/2024-25/FRG/93.

\bibliographystyle{IEEEtran}
\bibliography{qq}

\end{document}